\documentclass[preprint,amsmath,amssymb, aps, letterpaper, superscriptaddress]{revtex4}
\pdfoutput=1 
\usepackage{graphicx}% Include figure files
\usepackage{bm}% bold math

\newcommand{\be}{\begin{equation}}
\newcommand{\ee}{\end{equation}}
\newcommand{\bfig}{\begin{figure}}
\newcommand{\efig}{\end{figure}}
\newcommand{\incl}{\includegraphics}
\begin{document}

\title{Trajectory of Anomalous Hall Effect toward the Quantized State in a Ferromagnetic Topological Insulator}

\author{J. G. Checkelsky}
\altaffiliation[Present Address:]{ Department of Physics, Massachusetts Institute of Technology, Cambridge, MA 02139}
\affiliation{Department of Applied Physics and QPEC, University of Tokyo, Hongo, Tokyo 113-8656, Japan}

\author{R. Yoshimi}
\affiliation{Department of Applied Physics and QPEC, University of Tokyo, Hongo, Tokyo 113-8656, Japan}

\author{A. Tsukazaki}
\affiliation{Institute for Materials Research, Tohoku University, Sendai 980-8577, Japan}

\author{K. S. Takahashi}
\affiliation{RIKEN Center for Emergent Matter Science (CEMS), Wako 351-0198, Japan}

\author{Y. Kozuka}
\affiliation{Department of Applied Physics and QPEC, University of Tokyo, Hongo, Tokyo 113-8656, Japan}

\author{J. Falson}
\affiliation{Department of Applied Physics and QPEC, University of Tokyo, Hongo, Tokyo 113-8656, Japan}

\author{M. Kawasaki}
\affiliation{Department of Applied Physics and QPEC, University of Tokyo, Hongo, Tokyo 113-8656, Japan}
\affiliation{RIKEN Center for Emergent Matter Science (CEMS), Wako 351-0198, Japan}

\author{Y. Tokura}
\affiliation{Department of Applied Physics and QPEC, University of Tokyo, Hongo, Tokyo 113-8656, Japan}
\affiliation{RIKEN Center for Emergent Matter Science (CEMS), Wako 351-0198, Japan}

\date{\today}

%\end{abstract}

\maketitle                   % Produces the title

\textbf{Topological insulators are bulk electronic insulators which possess symmetry protected gapless modes on their surfaces \cite{Kane05, Bern06, Konig07, Fu07, Hsieh08}.  Breaking the symmetries that underlie the gapless nature of the surface modes is predicted to give rise to exotic new states of matter \cite{HasanRev}.  In particular, it has recently been predicted \cite{QAH1, QAH2} and shown \cite{QAHexp} that breaking of time reversal symmetry in the form of ferromagnetism can give rise to a gapped state characterized by a zero magnetic field quantized Hall response and dissipationless longitudinal transport known as the Quantum Anomalous Hall (QAH) state.  A key question that has thus far remained experimentally unexplored is the relationship of this new type of quantum Hall state with the previously known orbitally driven quantum Hall states.  Here, we show experimentally that a ferromagnetic topological insulator exhibiting the QAH state is well described by the global phase diagram of the quantum Hall effect.  By mapping the behavior of the conductivity tensor in the parameter space of temperature, magnetic field, and chemical potential in the vicinity of the QAH phase, we find evidence for quantum criticality and delocalization behavior \cite{Khm, Wei} that can quantitatively be described by the renormalization group properties of the quantum Hall ground state \cite{Pruis2, Dolan}.  This result demonstrates that the QAH state observed in ferromagnetic topological insulators can be understood within the context of the law of corresponding states which governs the quantum Hall state \cite{Kivelson91, Lutken}.  This suggests a roadmap for studying the QAH effect including transitions to possible adjacent topologically non-trivial states and a possible universality class for the QAH transition.}

The introduction of magnetism has proven to be an incisive method to experimentally access the exotic nature of topological insulators (TIs).  Early theoretical work predicted ordering of magnetic spins at the surface of TIs would open a gap at the Dirac point in the protected surface electronic spectrum \cite{Liu, Dima}.  This gap was observed by angle resolved photoemission spectroscopy (ARPES) experiments in both bulk \cite{ZXarpes} and surface \cite{HasanFM1} magnetically doped TIs.  Such bulk \cite{Chang2} and surface \cite{NatPhys} magnetic orderings were also observed to interact with the electronic transport of the surface modes.  The nature of the ground state within the electronic gap was predicted to be characterized by a zero field quantized Hall response (the QAH effect) in electronic transport \cite{QAH1} accessible under the condition of sufficient electronic localization \cite{QAH2}.  This line of inquiry culminated in the experimental observation of the QAH state in ferromagnetic TIs at low temperature \cite{QAHexp}.  With this discovery, significant interest has been focused on understanding the nature of the QAH state and its relation to conventional quantum Hall (QH) states.   

Here we establish an understanding of these symmetry broken TI surface modes within the general context of the stable ground states of two-dimensional (2D)  systems.  In the presence of disorder and absence of magnetic fields, it is known that as temperature $T$ goes to zero 2D systems do not support electrical conduction \cite{Abrahams}.  With the introduction of a magnetic field, there is an additional stable state characterized by vanishing longitudinal conductance and quantized Hall conductance $ne^2/h$ ($n$ is a non-zero integer, $e$ is the electronic charge, and $h$ is Planck's constant): the QH liquid \cite{Pruis1}.  2D systems subject to time reversal symmetry (TRS) breaking due to ferromagnetism rather than external magnetic field have also been shown theoretically to support stable insulating and QH liquid states \cite{Onada}.  We are thus led to consider the surface states of magnetically doped TIs as 2D transport systems in the presence of symmetry breaking ferromagnetism.  As we show below, in the presence of magnetic order these states can be successfully understood as stabilized QH liquids and are fully characterized by the universal delocalization response known in QH systems.

We have grown thin films of the topological insulator Cr$_{x}$(Bi$_{1-y}$Sb$_{y}$)$_{2-x}$Te$_{3}$ on semi-insulating InP (111) substrates using molecular beam epitaxy (see methods).  The films presented here have a thickness $t$ of approximately 8 nm estimated from the calibrated growth rate and a fixed Bi/Sb ratio set by $y=0.8$.  For this $y$ value, it has been shown by ARPES that the Dirac point of the surface states is isolated within the bulk electronic band gap \cite{Jiang}.  To demonstrate the effect of the magnetic dopant Cr, we compare results for growth of pristine ($x = 0$) and doped ($x=0.22$) films.  Fig. \ref{Fig1}(a) and (b) shows the X-ray diffraction pattern for the pristine and doped films, respectively; in each case all peaks can be identified with (0 0 0 $n$) diffraction of (Bi$_{1-y}$Sb$_{y}$)$_{2}$Te$_{3}$ or the InP $(n$ $n$ $n)$ peaks.  Fig. \ref{Fig1}(c) shows a detailed view of the (0 0 0 15) peak exhibiting Laue fringes consistent with our estimated $t$.  A compression of the $c$-axis lattice parameter is apparent in the Cr-doped film, reducing from $3.047$ nm to $3.026$ nm, suggestive of Cr replacing Bi.  The left and right inset of Fig. \ref{Fig1}(c) show the topography of the pristine and doped films, respectively, taken by atomic force microscopy.  The pristine film shows an almost atomically flat surface, i.e. it is comprised almost entirely of two quintuple layer (QL) levels separated by the QL thickness (1 nm), whereas the surface of the doped film shows approximately 3-QL levels.  From this comparison we conclude that the introduction of Cr increases disorder in the films, but the film nevertheless retains the expected crystalline structure and quality.   

We next compare the electronic transport properties of the films.  Fig. \ref{Fig1}(d) shows resistance $R$ as a function of temperature $T$ for both films.  Decreasing from $T=300$ K, the pristine film exhibits a non-metallic $R(T)$ to $T \approx$ 80 K, followed by metallic behavior to 10 K, and finally an upturn to 2 K.  As has been previously discussed, these regimes arise when the chemical potential $\mu$ resides in the bulk band gap of a TI such that at high temperature excitation of carriers in to the bulk bands dominates transport giving way to an intermediate temperature regime dominated by metallic surface conduction and finally quantum corrections to conduction at the lowest temperatures \cite{PRLgate}.  Upon doping with Cr, the most prominent difference observed is an enhancement in $R(T)$ peaking at a critical temperature $T_C = 45$ K.  As shown in Fig. \ref{Fig1}(e) and (f), this $T_C$ also corresponds to the onset of hysteresis in the longitudinal $R_{xx}(B)$ and transverse $R_{yx}(B)$ magnetotransport for the Cr doped films, suggesting that $T_C$ can be associated with the magnetic ordering temperature of the films.  The behavior in Fig. \ref{Fig1}(e) and (f) is that of a typical ferromagnetic metal, where $R_{xx}$ has a peak at the coercive field $H_C$ (here, $H_C= 0.16$ T at $T = 2$ K) and $R_{yx}$ is dominated by the anomalous Hall effect \cite{NagaosaRMP}.  Above $T_{C}$ we can estimate the electron mobility $\mu_{e}$ of the films; at $T = 80$ K we have $\mu_{e} \approx 270$ cm$^{2}$ V$^{-1}$ s$^{-1}$ (in comparable conditions the pristine films exhibit $\mu_{e}$ approaching 700 cm$^{2}$ V$^{-1}$ s$^{-1}$).  We note that the parameters $t$ and $x$ are optimized for this experiment by examining $R(T)$ and $R_{yx}(B)$ so as to retain metallic behavior with the largest anomalous Hall response (see Supplementary Materials).    

In order to have \textit{in situ} control of $\mu$, we patterned the films in to Hall bars with deposited Ti/Au top gates and Al$_2$O$_3$ gate dielectrics (see methods).  An image of a patterned device is shown as the left inset of Fig. \ref{Fig1}(d) along with a schematic representation of the vertical structure inset right.  
For the films here, we are able to reach charge neutrality ($\mu$ balanced between electron-like and hole-like) at a film-dependent top gate voltage $V_{T}$ that we hereafter refer to as $V_0$.  The results for both the pristine and doped films at $T = 2$ K and $B = 5$ T are shown in Fig. \ref{Fig1}(g) as a function of $V_T$ relative to $V_0$.  
For the pristine film, we see ambipolar behavior indicated by the sign change in $R_{yx}(V_T)$ at  $V_{T} = V_0$ and the simultaneous peak in $R_{xx}(V_T)$ \cite{PRLgate}. For the doped film, we also observe a peak in $R_{xx}(V_T)$, though it markedly less symmetric, along with a simultaneous variation in $R_{yx}(V_T)$. Here, we define $V_0$ at the peak in $R_{yx}(V_T)$ although no sign change occurs because of the large contribution of the anomalous Hall response. The observation of a maximum in the anomalous Hall response at charge neutrality is consistent with previously reported behavior \cite{Chang2}.

Cooling to lower temperatures, we observe behavior characteristic of the QAH effect.  In Fig. \ref{Fig2}, measurements of a doped film at $V_{T}=V_{0}$ and $T = 50$ mK are shown up to a magnetic field of $B= 14$ T.  $R_{yx}$ exhibits a value of 0.98 $\pm 0.003$ $h/e^{2}$ while the longitudinal resistance falls in increasing $B$, reaching values $< 0.03$ $h/e^2$ at $B=14 $ T.  The inset of Fig. \ref{Fig2} shows the response at low magnetic field, highlighting a remnant anomalous Hall resistance of 0.98 $h/e^2$.  These results are similar to those previously reported \cite{QAHexp}, though here a larger $B$ is required to suppress $R_{xx}$ as well as $T_{C}$ and $H_{C}$ discussed above.  It appears that while substantial $B$ is needed to support dissipationless longitudinal transport, the Hall response is robust even in vanishing $B$, with the slight reduction from $h/e^2$ likely due to remnant conduction channels \cite{NatPhys}.  

To examine transport behavior in the vicinity of the QAH state, we show the detailed evolution of $R_{xx}$ and $R_{yx}$ at $B=14 $ T at various temperatures as a function of $V_{T}$ in Fig. \ref{Fig3}(a) and (b), respectively.  A nearly dissipationless $R_{xx}$ is observed over a narrow $V_{T}$ range around $V_{0} \approx 3 $ V at $T = 50$ mK which quickly weakens with increasing $T$ in a metallic fashion.  There is a notable asymmetry for $R_{xx}$ on the electron and hole sides, but both sides become non-metallic at sufficiently large $|V_{T}|$.  For $R_{yx}$ we observe a plateau over a slightly larger $V_{T}$ domain, which recedes with both increasing $T$ or for $V_{T}$ significantly away from $V_{0}$.  Mapping these curves in to conductivities produces a systematic view of this evolution.  In Fig. \ref{Fig3}(c)-(e) we make a parametric plot of $(\sigma_{xy}(V_{T}), \sigma_{xx}({V_{T}}))$, $V_{T}$ being the parameter, at various $B$ for $T = 700$ mK, 200 mK, and 50 mK, respectively.  At the lowest $T$ (Fig. \ref{Fig3}(e)), the evolution from low $V_{T}$ begins from large $\sigma_{xx}$ toward an apparent critical point at $(e^{2}/h,0)$, though with a minor deviation consistent with $R_{yx}$ being slightly below $h/e^2$ and the finite remnant  $R_{xx}$.  Interestingly, as $V_{T}$ passes through $V_{0}$ (in the vicinity of the critical point) and further increases, curves at all $B$ collapse on a single line nearly described by a semicircle of radius $e^{2} / 2h$ centered at $(e^{2}/2h,0)$, shown as a dashed line in Fig. \ref{Fig3}(e).  This recalls the semicircular law derived to describe the transition between adjacent QH states or to the insulator \cite{Circ1, Shay}.  The breakdown of this behavior appears to connect with the electron-hole asymmetry discussed above; we hypothesize that the asymmetry of the electronic structure with bulk valence band states being in closer proximity to the Dirac point may be the origin of this.  As more readily expected from theory, the deviation from the semicircular behavior increases with increasing $T$ (Fig. \ref{Fig3}(c) and (d)).  This is a preliminary connection of the present QAH state to known QH behavior.  

We can further probe the connection of the present system to QH states by examining the detailed temperature dependence of the conductivity tensor.  In Fig. \ref{Fig4}(a) $R_{xx}(T)$ is shown for several $V_{T}$ for $B=0$ (after application of large $B$ to saturate the magnetization $M$).  A crossover from non-metallic to metallic behavior across $h/e^{2}$ is observed as $V_{T}$ crosses +4 V.  In Fig. \ref{Fig4}(b) we show $R_{yx}(T)$ measured under the same conditions, where we observe a change from being increasing to decreasing $R_{yx}(T)$ with decreasing $T$ as $V_{T}$ is increased across +4 V.  This apparently complex crossover between two phases (one insulating and one metallic) is resolved by viewing these results from the perspective of the renormalization group (RG) properties of QH states.  

As noted above, in the absence of a TRS breaking magnetic field, 2D systems flow toward the insulating ground state as $T$ approaches zero (or equivalently as the system size $L$ diverges), a notion based on single parameter scaling analysis \cite{Abrahams}.  In the presence of broken TRS, however, the scaling function (the so-called $\beta$ function) involves two parameters.  Theoretically, the behavior of the conductivity tensor based on these $\beta$ functions under the condition of diverging $L$ under RG flow is suggested to be characterized by flow lines dictated by stable and unstable critical points in the parameter space of $\sigma_{xx}$ and $\sigma_{xy}$ \cite{Khm}.  The law of corresponding states for QH systems, the rules that govern the symmetry relations under which QH states are identical and thus dictate the phase diagram of QH systems \cite{Kivelson91}, plays a key role in allowing us to quantitatively draw a comparison between experiment and this theoretical description.

The symmetries embodied by the law of corresponding states have been shown to impose a symmetry corresponding to the $\Gamma_{0}(2)$ modular subgroup on the associated conductivity tensor \cite{Lutken}.  By writing the complex conductivity $\sigma = i\sigma_{xx}+\sigma_{xy}$, one can describe the behavior of $\sigma$ in the upper half of the complex plane under RG flow.  Here, we employ work motivated by the proposed duality of QH systems and $\mathcal{N}=2$ supersymmetric Yang-Mills theory rooted in their common $\Gamma_{0}(2)$ symmetry \cite{Dolan}.  In analogy to RG flow studied in the latter \cite{Witten}, it is postulated that the relevant $\beta$ functions are complex analytical functions of $\sigma$, from which the following function $f(x)$ can be derived
\be
f(x)=-\frac{\vartheta^{4}_{3}(x)\vartheta^{4}_{4}(x)}{\vartheta^{8}_{2}(x)},
\label{f}
\ee
where $\vartheta_{i}(x)$ are Jacobi theta functions of the $i$th kind, such that for $x = e^{i\pi\sigma}$ Eq. \ref{f} has a constant complex phase $\phi$ along RG flows, viz. $\textrm{arg}$ $f = \phi$ \cite{Dolan}.  We can then construct the RG flow diagram for the QH and insulating states by plotting the contours of arg $f$ in $\sigma$.  Several contours in the phase space covering the insulator and filling $\nu=1$ QH state are shown in Fig. \ref{Fig4}(c), with arrows indicating the direction of flow with increasing length scale.  In the vicinity of (0,0) and ($e^2/h,0$) there are two stable fixed points (denoted by $\odot$) corresponding to insulating and QH ground states, respectively.  The unstable point ($\otimes$) indicates the transition regime or delocalization regime where extended states exist.  These features reproduce the flows predicted by other methods  \cite{Pruis2, Khm} and characterize the phase transitions of the 2D electron system with broken TRS.  We note that setting $\phi=0$ recovers the condition for the semicircular law discussed above, which is not surprising as $\Gamma_{0}(2)$ symmetry serves as the basis for an alternate derivation \cite{Circ2}.  

Turning to comparison with our experimental results, we plot $(\sigma_{xy}(V_{T}), \sigma_{xx}(V_{T}))$ for $B=0$ at various $V_{T}$ in Fig. \ref{Fig4}(d).  As $T$ is decreased (corresponding to increasing $L$), $(\sigma_{xy}(V_{T}), \sigma_{xx}(V_{T}))$ appears to flow to one of two stable fixed points at $(0,0)$ and $(e^{2}/h,0)$, with the pattern suggesting an unstable fixed point in the vicinity of $e^2/h(0.5, 0.55)$.  This unstable fixed point coincides closely with that expected for the delocalization transition of extended states producing universal singularities \cite{Pruisken}.  This behavior thus qualitatively reproduces the characteristics of renormalization group flow toward the insulating and QH ground states in Fig. \ref{Fig4}(c).  Quantitatively, we find that Eq. \ref{f} describes flow lines that capture our experimental result with reasonably high accuracy and precision, shown as the thin black curves in Fig. \ref{Fig4}(d).  Previous experimental work has captured aspects of this flow in heterojunction QH systems \cite{Wei}; this is clear evidence that the QAH state obeys the same symmetry rules.  This flow is controllable with $V_{T}$ even with $B=0$, suggesting that spontaneous $M$ in the QAH can drive identical behavior to $B$ in QH systems (similar behavior is observed in field, see supplementary materials).

Comparing to previous systems in which RG flow has been less successful in the regime of classical percolation \cite{Wei}, one conclusion here is that quantum localization plays a key role in determining the observed behavior.  The agreement between the behaviors of the QAH effect and QH effect in terms of RG flow may not be entirely surprising given that both the anomalous Hall effect and QH effect are described by field theories that are topological in origin \cite{NagaosaRMP, Pruis2}.  This observation motivates a number of directions for future study.  One extension of this analogy is to the larger global phase diagram of QH states for the QAH effect, which would  predict that the $\nu = 1$ QAH state would be connected continuously to other integer and fractional QAH states and that other exotic insulating states may be in close proximity \cite{Kivelson91}.  If there are indeed neighboring states this suggests that the corresponding phase transitions would belong to the same universality class in the Anderson localization regime and would be of great interest for further study.  As has been observed previously \cite{Wei}, we would also predict this scaling would break down at elevated temperatures.  We expect this framework for understanding the QAH effect will guide the discover of new quantum states and phase transitions in the study of TIs and TRS breaking.

\section*{Methods}

Thin films are grown by molecular beam epitaxy on epi-ready semi-insulating InP (111) at a substrate temperature of 260$^{\circ}$ C after annealing at 380$^{\circ}$ C.  The beam flux of Te relative to Bi, Sb, and Cr is kept at a ratio of of 20:1 to suppress Te vacancies.  The elemental compositions of Bi, Sb, and Cr are estimated from the relative beam flux pressures checked before and after deposition.  The growth rate of the films was approximately 0.2 QL / min. Transport measurements for $T > 2$ K are performed on unpatterned films with an excitation current of 1 $\mu$A via electrical contacts made with Au wires and Ag paint in a commercial variable temperature cryostat.  To correct for contact misalignment, the data in Fig. 1(e) and (f) are field symmetrized (unsymmetrized data is shown in Fig. S7).  For low temperature measurements, films are capped \textit{ex situ} after growth with Al$_{2}$O$_{3}$ deposited by atomic layer deposition (performed at $T <$ 120 $^{\circ}$C to a thickness of 25 nm) before subsequent processing by standard photolithography techniques and Ar ion milling (the etching is done with a brief exposure of less than 60 s).  Electrical contacts and the top gate are made with e-beam evaporation of Ti/Au.  Care is taken to use low $T < 150$ $^{\circ}$C processing to avoid degradation of the films.  $R_{xx}$ and $R_{yx}$ reported here are the longitudinal (sheet) resistance and transverse (Hall) resistance, respectively.  Measurements on devices are performed at low frequency (1-3 Hz) with lock-in amplifiers and voltage and current pre-amplifiers with an excitation current limited to 1 nA.  We use a commercial dilution refrigerator equipped with a superconducting magnet.  For device measurements we do not perform field symmetrization.

\vspace{10 mm}

\textbf{Acknowledgments} We are grateful to N. Nagaosa, B.-J. Yang, and A. F. Young for fruitful discussions and M. Nakano, T. Hatano, S. Shimizu, M. Kubota, and S. Ono for technical support. This research is supported by the Japan Society for the Promotion of Science (JSPS) through the ``Funding Program for World Leading Innovative R\&D on Science and Technology (FIRST Program),'' initiated by the Council for Science and Technology Policy (CSTP), and by JPSP Grant-in-Aid for Scientific Research, Nos. 24224009, 24226002, and 25871133.  This work was carried out by joint research of the Cryogenic Research Center, the University of Tokyo.

\textbf{Author Contributions} J.G.C. and R.Y. grew and characterized the films.  J.G.C., Y.K., and J.F. performed the low temperature measurements.  J.G.C. analyzed the data and wrote the letter with contributions from all authors.  A.T., K.S.T., M.K., and Y.T. contributed to discussion of the results and guided the project. Y.T. conceived and coordinated the project.   

\textbf{Author Information} The authors declare no competing financial interests.  Correspondence and requests for materials should be addressed to J.G.C.

\pagebreak

%%%%%%%%%%%%%%%%%%%%%%%%%%%%%%%%%%%%%%%%
%%%%%%%%%%%%%%%%%%%%%%%%%%%%%%%%%%%%%%%%
%%%%%%%%%%%%%%%%%%%%%%%%%%%%%%%%%%%%%%%%
%%%%%%%%%%%%%%%%%%%%%%%%%%%%%%%%%%%%%%%% FIGURE 1
\bfig[htb]            % Fig 1
%\incl[width=8.65cm]{Figver003-01.eps}
\incl[width=13cm]{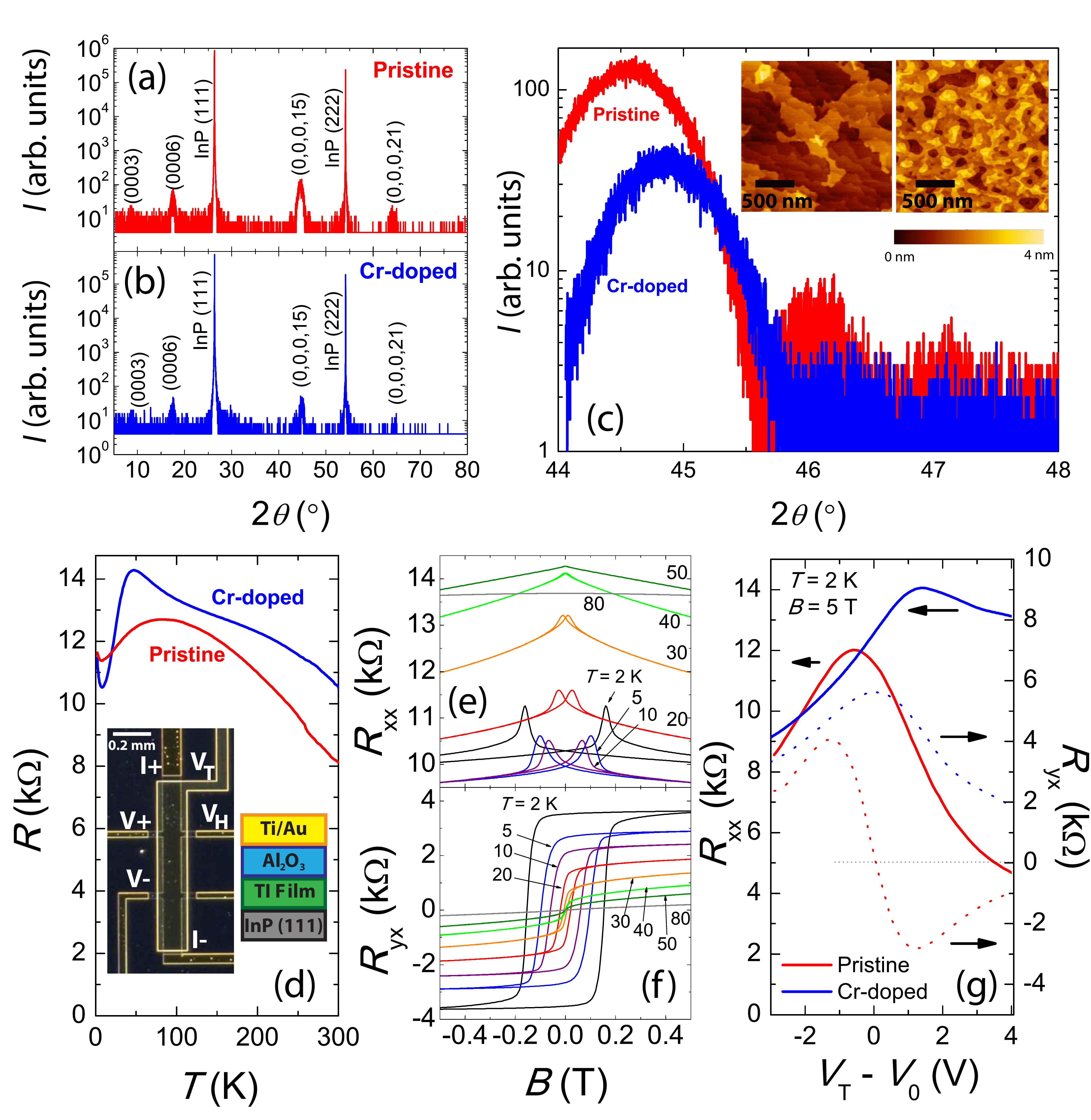}
\caption{\label{Fig1}
X-ray diffraction pattern for Cr$_{x}$(Bi$_{1-y}$Sb$_{y}$)$_{2-x}$Te$_{3}$ with $y=0.8$ for \textbf{(a)} the pristine ($x=0$) and \textbf{(b)} optimally magnetically doped ($x=0.22$) cases.  \textbf{(c)} Detailed view of the (0 0 0 15) peak.  The inset shows 2 $\mu$m $\times$ 2 $\mu$m atomic force microscope topography of the films.  \textbf{(d)} Resistance $R$ as a function of temperature $T$ for the films.  The inset shows a lithographically defined device used in this study along with a schematic of the electrostatic gate structure.  \textbf{(e)}  Longitudinal resistance $R_{xx}$ and \textbf{(f)} transverse resistance $R_{yx}$ of the Cr-doped film as a function of magnetic field $B$ showing hysteresis associated with onset of ferromagnetic ordering near $T=$ 40 K.  \textbf{(g)} Gate voltage $V_{T}-V_{0}$ (relative to the zero gap voltage) dependence of pristine and Cr-doped films at $T = 2 $ K and $B = 5$ T showing ambipolar behavior.
}
\efig

\pagebreak

%%%%%%%%%%%%%%%%%%%%%%%%%%%%%%%%%%%%%%%%
%%%%%%%%%%%%%%%%%%%%%%%%%%%%%%%%%%%%%%%%
%%%%%%%%%%%%%%%%%%%%%%%%%%%%%%%%%%%%%%%%
%%%%%%%%%%%%%%%%%%%%%%%%%%%%%%%%%%%%%%%% FIGURE 2
\bfig[htb]            % Fig 2
\incl[width=13cm]{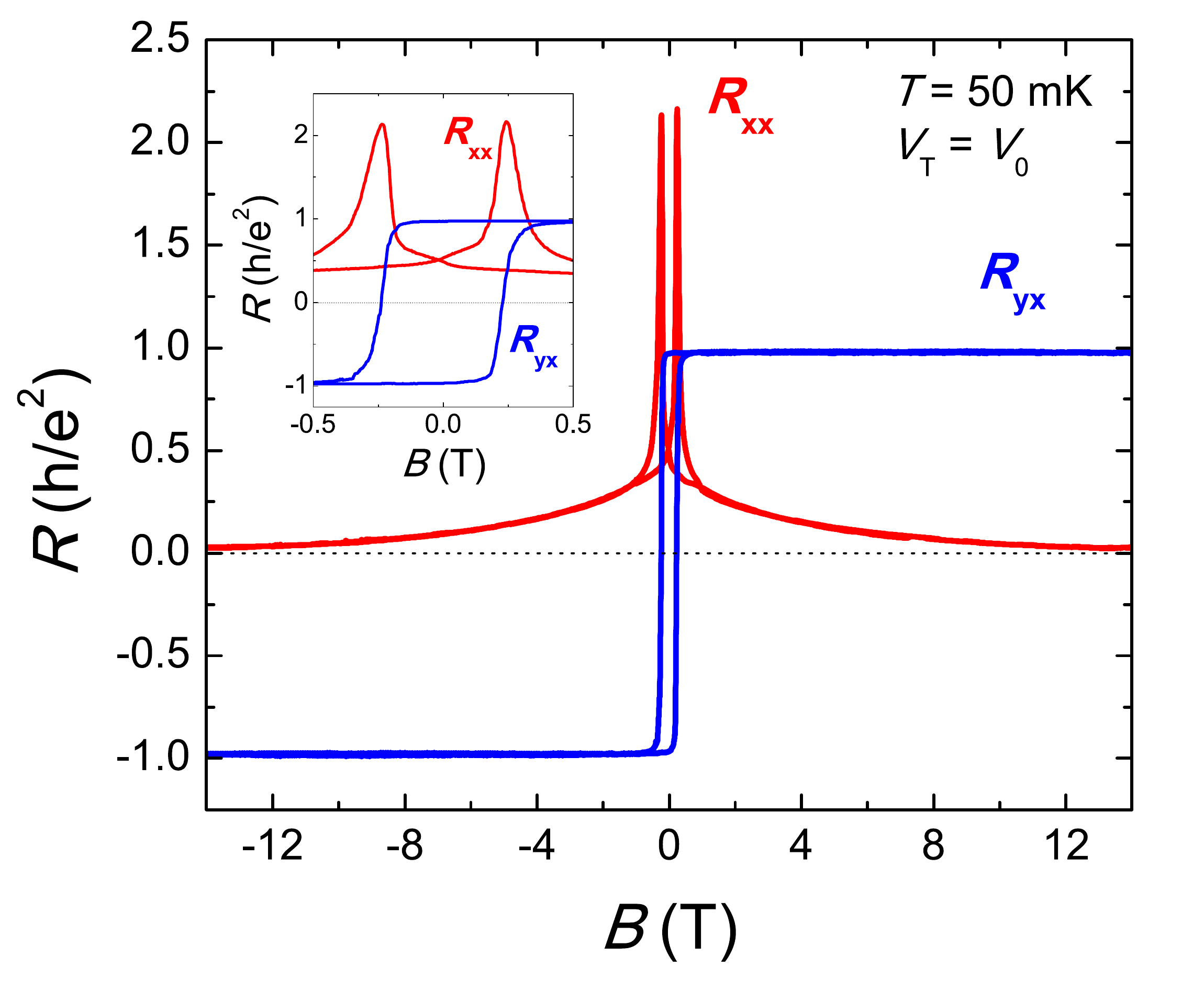}
\caption{\label{Fig2}
Longitudinal resistance $R_{xx}$ and transverse resistance $R_{yx}$ of Cr-doped film as a function of magnetic field $B$ at $T = 50$ mK and optimally tuned gate voltage $V_{0}$.  $R_{yx}$ exhibits a value of 0.98 $\pm 0.003$ $h/e^{2}$ while the longitudinal resistance falls in increasing $B$, reaching values $< 0.03$ $h/e^2$ at $B=14 $ T.  The inset shows the response at low magnetic field.
}
\efig

\pagebreak

%%%%%%%%%%%%%%%%%%%%%%%%%%%%%%%%%%%%%%%%
%%%%%%%%%%%%%%%%%%%%%%%%%%%%%%%%%%%%%%%%
%%%%%%%%%%%%%%%%%%%%%%%%%%%%%%%%%%%%%%%%
%%%%%%%%%%%%%%%%%%%%%%%%%%%%%%%%%%%%%%%% FIGURE 3
\bfig[htb]            % Fig 3
\incl[width=13cm]{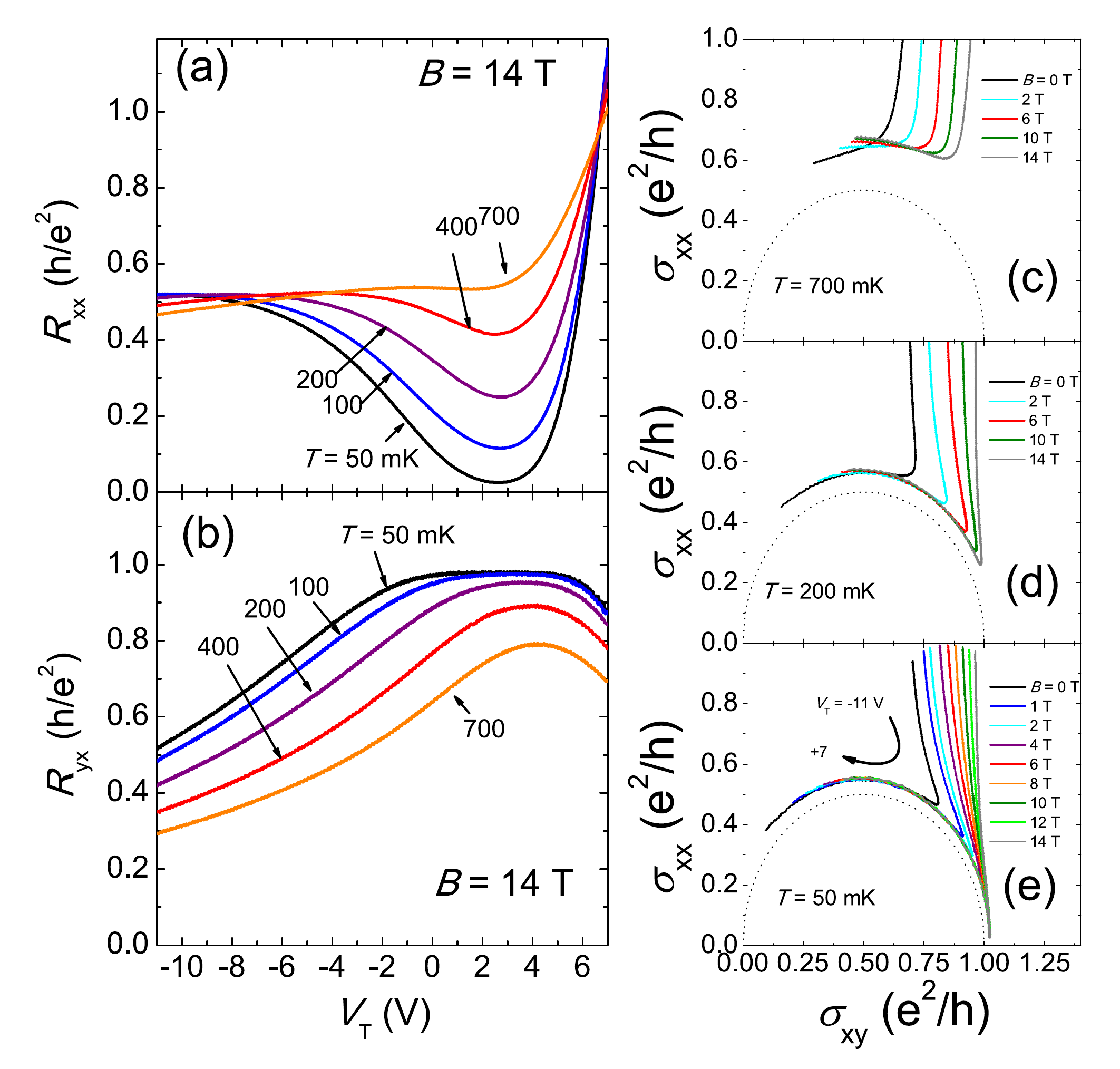}
\caption{\label{Fig3}
\textbf{(a)} Top gate voltage $V_{T}$ dependence of longitudinal resistance $R_{xx}$ at $B= 14$ T.  \textbf{(b)}  $V_{T}$ dependence of transverse resistance $R_{yx}$ at $B= 14$ T.  \textbf{(c)-(e)} $(\sigma_{xy}(V_{T}), \sigma_{xx}(V_{T}))$ for various $B$ at $T = 700$ mK, 200 mK, and 50 mK.  The direction of flow is clockwise for negative to positive gate voltage, as depicted by the arrow.  A semicircle of radius $e^{2}/2h$ centered at $(e^{2}/2h, 0)$ is shown as a dashed line in each panel.
}
\efig

\pagebreak

%%%%%%%%%%%%%%%%%%%%%%%%%%%%%%%%%%%%%%%%
%%%%%%%%%%%%%%%%%%%%%%%%%%%%%%%%%%%%%%%%
%%%%%%%%%%%%%%%%%%%%%%%%%%%%%%%%%%%%%%%%
%%%%%%%%%%%%%%%%%%%%%%%%%%%%%%%%%%%%%%%% FIGURE 4
\bfig[htb]            % Fig 4
\incl[width=16cm]{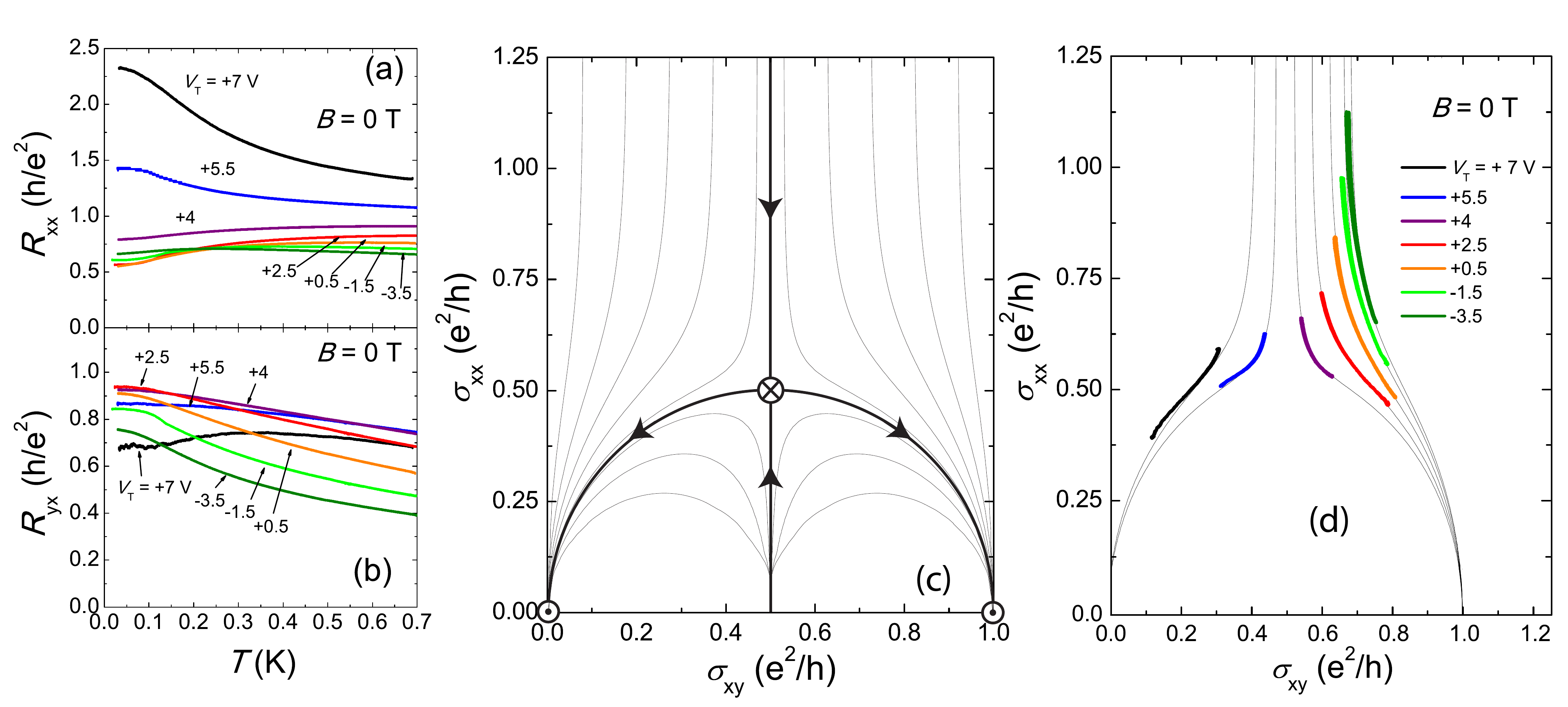}
\caption{\label{Fig4}
\textbf{(a)} Temperature $T$ dependence of longitudinal resistance $R_{xx}$ at $B=0$ (after magnetic training) for various top gate voltages $V_{T}$.  \textbf{(b)} $T$ dependence of transverse resistance $R_{yx}$ at $B=0$ (after magnetic training) for various $V_{T}$.  \textbf{(c)} Calculation of Renormalization Group (RG) flow for a 2D electron system with insulating and quantum Hall stable points.  \textbf{(d)} Flow in $(\sigma_{xy}(V_{T}), \sigma_{xx}(V_{T}))$ observed at $B=0$ at various $V_{T}$.  Upon decreasing $T$, the systems evolves toward lower $\sigma_{xx}$.  The thin lines are RG flow lines calculated as discussed in the text.
}
\efig

\end{document}